\documentclass[showpacs,preprintnumbers,amsmath,amssymb]{revtex4}

\usepackage{graphicx}
\usepackage{dcolumn}
\usepackage{bm}

\begin{document}

\preprint{APS/123-QED}

\title{Laser Stripping via a Broad Stark State for a High-Intensity Proton Ring}

\author{Isao Yamane}
 \email{isao.yamane@kek.jp}
\author{Toshikazu Suzuki}
\author{Ken Takayama}
\affiliation{KEK, High Energy Accelerator Research Organization, Oho 1-1, Tsukuba, Ibaraki, 305-0801 Japan
}

\date{\today}

\begin{abstract}
A new type of charge-exchange injection scheme for high-intensity proton rings that use a laser beam and an undulator magnet is proposed. The elementary stripping process is resonant photoionization via a broad Stark state formed by the Stark effect at an energy level near the peak of the potential barrier. The method used to estimate the parameters of the necessary laser beam and magnetic field is described. As an application to an actual accelerator, a laser stripping system using a high-finesse Fabry-Perot resonator and an undulator is proposed. An estimation of the photon loss due to the pumping-up of H$^0$ atoms, tunability of the system and the emittance growth due to stripping in the undulator magnet is also given.
\end{abstract}

\pacs{Valid PACS appear here}
\maketitle

\section{\label{sec:level1}Introduction}

Thin foils of carbon or alminum oxide are very convenient for stripping media for H$^-$ charge-exchange injection into a proton ring. However, two serious problems come to the fore for such MW-class proton rings as drivers of the next-generation neutron sources or neutrino factories. One is a serious degradation of the foil performance due to heating to an extremely high temperature by the energy deposit of traversing protons. The other is activation of the accelerator components by high-energy protons scattered by foil atoms \cite{ref1}.

In order to avoid these problems, stripping methods without foils have been studied from several years ago. Those methods usually combine Lorentz stripping and the excitation of H$^0$ atoms by laser pumping. Here, Lorentz stripping means stripping an electron from an H$^-$ ion or an H$^0$ atom using the Lorentz force that acts on charged particles with a velocity in a magnetic field \cite{ref2, ref3, ref4}.

With respect to the Lorentz stripping of an H$^-$ beam, we well know about the experiences of the PSR (Proton Storage Ring) at Los Alamos National Laboratory \cite{ref5}. A two-step H$^0$ injection scheme, that used neutralization of H$^-$ beams by Lorentz stripping for the first step, has finally been converted to direct H$^-$ injection without Lorentz stripping, because the controllability of the beam size at the foil was very low for H$^0$ beams. Since the beam size of H$^0$ beams could not be reduced sufficiently, neither the area of the stripping foil that protons captured in the ring hit, nor the foil-hit number of the captured proton beam could be sufficiently reduced. However, there is a suggestion that the reason for the low controllability of the H$^0$ beam size at the foil is the foil location in PSR, and that perfect matching of the H$^0$ beam to the ring optics is possible by choosing the foil location \cite{ref6}. In any case, these problems will not take place if a stripping foil is not used. In addition, when the energy of the H$^-$ beam becomes higher, the dissociation probability of H$^-$ ions by the Lorentz force becomes higher, and the angular divergence of the H$^0$ beam in a stripping magnet using a steep magnetic field gradient becomes smaller. Therefore, if the optics of the H$^-$ and H$^0$ beams is carefully designed, Lorentz stripping is considered to be sufficiently useful, especially for higher energy H$^-$ beams.

With respect to the stripping of H$^0$ beams, we know that a beam of excited H$^0$ atoms with a principal quantum number of no less than 3 may be stripped by Lorentz stripping in the same manner as a H$^-$ beam \cite{ref1}. H$^0$ atoms formed by Lorentz stripping from H$^-$ ions are left in the ground state. If H$^0$ atoms of the beam can be excited to a state with a principal quantum number of no less than 3 by laser pumping, a stripping scheme without foils will be possible. Thus, a three-step stripping scheme was studied first. The first step is Lorentz stripping of a H$^-$ beam to a H$^0$ beam by a stripping magnet. The second step is pumping-up of the H$^0$ beam to an excited state with the principal quantum number being no less than 3, using a laser beam in a straight section of the ring. Finally, the third step is again Lorentz stripping of the excited H$^0$ beam by another stripping magnet. A variety of three-step stripping schemes were reported \cite{ref1, ref7, ref8}. The major difference of these schemes are in the second step, where we must take care of the Doppler broadening of the transition frequency distribution due to the momentum spread of the H$^0$ beam.

It was known from an early stage of studies on the laser stripping that there are two problems for the second step of this scheme. One problem is that, since the transition frequency shifts by the Doppler effect, the transition frequency distribution spreads to as broad as $10^{13}$Hz, according to the momentum spread, $\Delta p / p$, of the H$^0$ beam, which is typically $10^{-3}$. We denote the frequency by the angular frequency, $\omega \equiv 2 \pi \nu$ , except when otherwise mentioned. The line width of the laser should be narrower than 10 MHz in order to secure a coherence time necessary to cover the flight time of light from the laser source to the end of the interaction region, which is typically on the order of 10 ns. Therefore, we need to take some measure to cover the spread of transition frequency as broad as $10^{13}$ Hz using a laser beam having a line width as narrow as 10 MHz. The other problem is that if we do not take any measure, the rate of H$^0$ atoms that are excited in one collision between H$^0$ and the laser beam is at most one half. This is because the probability of the absorption and the stimulated emission of a photon by a H$^0$ atom are the same, and the population of the excited state saturates to one half. Therefore, in order to achieve an efficiency near to 100\%, collisions between the H$^0$ beam and the laser must be repeated 4 or 5 times.

When H$^0$ atoms are placed in an external electric field, a potential barrier is formed due to the superposition of the Coulomb potential of the proton and the external electric field. The energy levels of the H$^0$ atoms shift due to the Stark effect \cite{ref9}. As the electric field is increased, the peak energy of the potential barrier becomes lower. And as the energy level comes near to the peak of the potential barrier, the lifetime of the energy level due to field ionization becomes shorter, and the level width becomes broader. Thus, the optical spectral lines fade out when the field ionization lifetime becomes comparable to the lifetime against radiation. However, the energy levels of the H$^0$ atoms are well defined at energies near to the peak of the effective potential barrier \cite{ref10}. The reciprocal lifetimes of those energy levels become as high as $10^{12} \sim 10^{13}$ sec$^{-1}$ and, accordingly, the widths may be as broad as $10^{13}$ Hz \cite{ref11}. Broadening of the level width with an increase of the electric field was experimentally observed using 800 MeV H$^-$ beams at the Clinton P. Anderson Meson Physics Facility \cite{ref12}.

The widths of these levels are as broad as the spread of the transition-frequency distribution, due to the momentum spread of the H$^0$ beam, and are controllable by varying the external electric field. Thus these levels were first used as a measure against the first problem mentioned above. However, these broad levels soon also turned out to be helpful as a measure against the second problem. Because excited H$^0$ atoms ionize within a very short lifetime, the population of excited H$^0$ atoms cannot increase, and therefore pumping down to the ground state by stimulated emission is negligible. Thus, the ionization efficiency is expected to be 100\%. The level broadened by the Stark effect as broad as the spread of the transition-frequency distribution due to the momentum spread of the H$^0$ beam is the essential element of the stripping scheme described here. Therefore, for the sake of convenience, it is called the broad Stark state.

\section{\label{sec:level2} Doppler Shift of the Transition Frequency}

\begin{figure}
\includegraphics{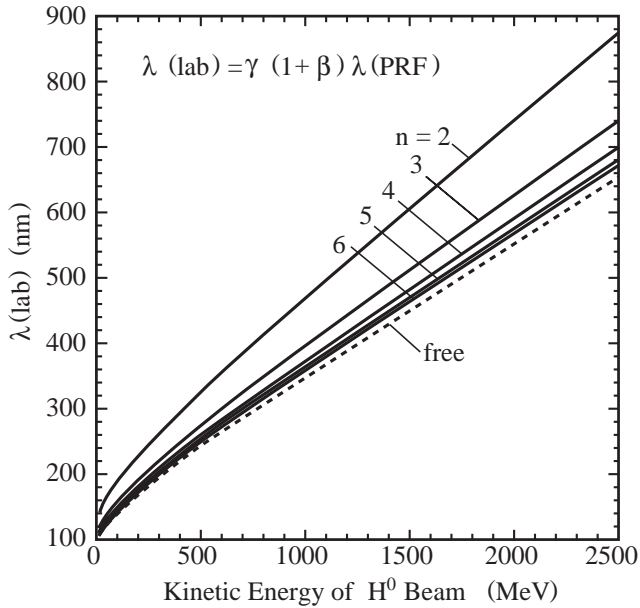}
\caption{Doppler shift of the transition wavelength for a H$^0$ beam. $\lambda$(lab) and $\lambda$(PRF) are the laser wavelengths in the laboratory and the particle rest frame, respectively. {\bf n} indicates the principal quantum number of the hydrogen atom. The dotted line indicates the transition wavelength between the ground state and the state of zero binding energy, and free means the electron is in the {\bf free} state.}
\label{fig1}
\end{figure}

The frequency ($\omega$) of laser in the laboratory frame is shifted to $\omega _0$ in the particle rest frame by the Doppler effect, and $\omega _0$ is given by
\begin{equation}
\omega _0 = \gamma \left( 1 + \beta \cos \alpha \right) \omega,
\label{eq1}
\end{equation}
where $\beta$ and $\gamma$ are relativistic parameters of the particle, and $\alpha$ is the angle between the velocity of the particle and the laser beam in the laboratory frame. Accordingly, the wavelength shifts with the kinetic energy of the H$^0$ beam, as shown in Fig.~\ref{fig1}. In the particle rest frame of an H$^0$ atom, the transition wavelength between the ground state and the state with a principal quantum number 3 is 103nm, and a laser with this wavelength is in the ultraviolet region. However, the transition wavelength is in the color region when the H$^0$ beam energy becomes several hundreds MeV or higher. Most of the beam energies of H$^-$ charge-exchange injection into high-intensity proton rings are in this energy region. Generally speaking, lasers in the color region are easier to obtain and handle than those in the ultraviolet region. This is a big advantage of the Doppler effect.

Because the momentum spread of the H$^0$ beam is the velocity spread of particles in the beam, the transition frequencies of the individual particles spread due to the Doppler effect. The spread of the transition-frequency distribution in the particle rest frame ($\Delta \omega _0$) is given by
\begin{equation}
\Delta \omega _0 = \omega _0 \beta \left( \frac{\Delta p}{p} \right),
\label{eq2}
\end{equation}
where $\omega _0$ is the transition frequency in the particle rest frame and $\Delta p / p$ is the momentum spread of the H$^0$ beam in the laboratory frame. For example, when the momentum spread of a 1 GeV H$^0$ atom is $10^{-3}$, as a typical value, $\Delta \omega _0 = 1.6 \times 10^{13}$ Hz for the transition frequency between the ground state and the state with a principal quantum number of 3.

\section{\label{sec:level3} Laser Stripping via a Broad Stark State}

When a particle, with relativistic parameters of $\beta$ and $\gamma$, moves in a magnetic field $B$ in the laboratory frame, the particle is affected in the particle rest frame by an electric field ($E$) given by 
\begin{equation}
E = \beta \gamma c B.
\label{eq3}
\end{equation}

\begin{figure}
\includegraphics{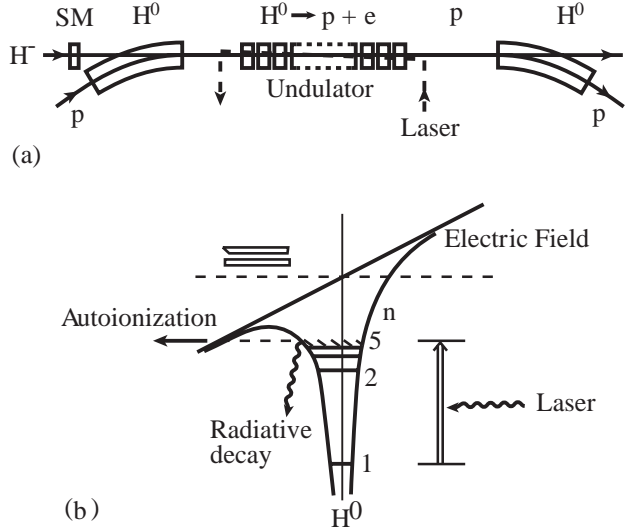}
\caption{(a) New type of the H$^-$ charge-exchange injection system using an undulator and a laser pumping system. SM indicates a stripping magnet. H$^0$ atoms formed by SM drift into the magnetic field of the undulator. Then, they are pumped up to a broad Stark state and decay to a proton and an electron. (b) Conceptual drawing of the stripping process. An aimed Stark state of H$^0$ (here the state with {\bf n} $= 5$) is broadened as much as the spread of the transition frequency due to the momentum spread of the H$^0$ beam by adjusting the undulator magnetic field so that the aimed comes near to the peak of the effective potential barrier. Then, H$^0$ atoms are pumped up to the broad Stark state by a narrow-band laser. Formed excited H$^0$ atoms immediately decay through autoionization.}
\label{fig2}
\end{figure}

This electric field causes the particle to experience the Stark effect. Therefore, by adjusting $B$ so that the Stark level comes near to the peak of the potential barrier, a broad Stark state can be formed. Combining such a broad Stark state and laser pumping, a new type of H$^-$ charge-exchange injection system, like Fig.~\ref{fig2}(a), becomes possible.

In this injection system, a H$^-$ beam is firstly converted to a H$^0$ beam by a stripping magnet. Next, the H$^0$ beam is conducted into a magnetic field of an undulator, where a broad Stark state of H$^0$ atoms is prepared. Then, the H$^0$ atoms are pumped up from the ground state to the prepared broad Stark state by a narrow-band laser beam. Excited H$^0$ atoms immediately dissociate and form a proton beam. Finally, the proton beam is captured by the ring.

The stripping process of this scheme is conceptually shown in Fig.~\ref{fig2}(b). When H$^0$ atoms come into the magnetic field of the undulator, the potential of the H$^0$ atom is deformed to form a barrier by the electric field of the Lorentz force. The magnetic field of the undulator is adjusted so that the aimed Stark state comes near to the peak of the potential barrier, and is broadened as broad as the spread of the transition-frequency distribution due to the momentum spread of the H$^0$ beam. Because the level width of the state is sufficiently broad, a narrow-band laser with the same frequency as the central transition frequency can interact with all H$^0$ atoms of the beam, although the distribution of the transition frequency is very broad due to the momentum spread. Then, H$^0$ atoms are pumped up to the broad Stark state by a narrow-band laser. A narrow-band laser is necessary to secure the coherence time that should cover the flight time of light from the laser source to the end of the interaction region. Excited H$^0$ atoms immediately ionize through auto-ionization, because the ionization probability of the state is far larger than the probability of the stimulated emission of photons. Therefore, the rate of pumping down to the ground state is negligible, and the reaction proceeds one-directionally from pumping up to auto-ionization. Competition between pumping up and down does not occur. Thus, the stripping efficiency is expected to reach 100\%. the ionization cross section is given as a product of the cross section of the transition from the ground state to the excited state and the ionization rate of the excited state. In this case, the ionization rate is almost one, because the ionization probability of the excited state is very high and the ionization cross section is almost equal to the transition cross section.

We now estimate the parameters of the laser and the undulator necessary to realize a stripping system for a H$^0$ beam. With respect to the Stark state of an H$^0$ atom in an external electric field, the variables describing the quantum state are separated in parabolic coordinates \cite{ref9}. An atomic state is completely defined by a set of quantum numbers ($n, k_1, k_2, m$). The quantum number $n(m)$ is the principal (magnetic) quantum number, $k_1$ is the quantum number of a partial wave function in a partial potential that becomes infinity at infinite distance, and $k_2$ is that in a partial potential that has a barrier and becomes negative infinity at infinite distance. Also, there is the relation $n = k_1 + k_2 + |m|+ 1$.

\begin{figure}
\includegraphics{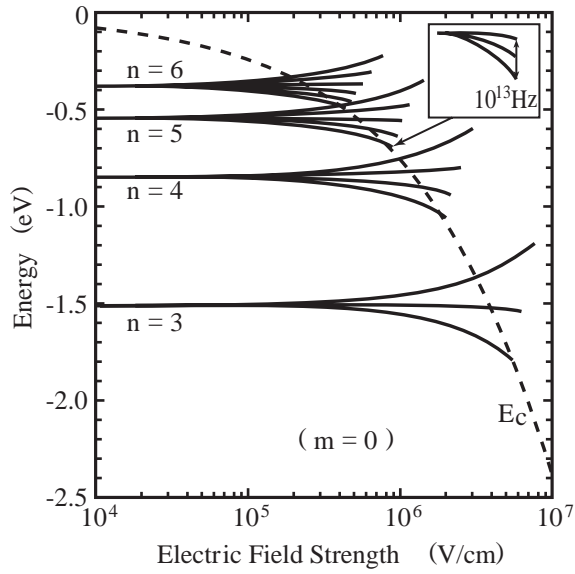}
\caption{Electric field dependence of the Stark levels. {\bf n} ({\bf m}) is the principal (magnetic) quantum number. {\bf E}$_{\mbox{\bf c}}$ is the energy of the peak of the potential barrier. Every level is terminated at the point where the level width is $10^{13}$ Hz, because the level width is rapidly broadened as {\bf E}$_{\mbox{\bf c}}$ comes near the level, and the level finally disappears after it exceeds {\bf E}$_{\mbox{\bf c}}$.}
\label{fig3}
\end{figure}

The electric field dependence of the energy level of the H$^0$ atom Stark state with a set of quantum numbers ($n, k_1, k_2, m$) is given by the following formula \cite{ref9}:
\begin{equation}
W=-{1 \over {2n^2}}+{3 \over 2}n\left( {k_1-k_2} \right)E-{1 \over {16}}n^4\left\{ {17n^2-3\left( {k_1-k_2} \right)^2-9m^2+19} \right\}E^2,
\label{eq4}
\end{equation}
where $W$ is the energy in atomic unit (a. u.=27.21eV) and $E$ is the strength of the external electric field in atomic unit (a. u.$=5.142 \times 10^9$V/cm). In Fig.~\ref{fig3}, the dependence of the energy level upon the external electric field is shown for several states of $n$ from 3 to 6 with $k_1 = 0$, $k_2 = n-1$, and $m = 0$. $E_c$ indicates the energy of the peak of the potential barrier. $E_c$ decreases with the electric field strength. When $E_c$ comes near a level, the decay probability of the level due to the tunneling effect increases and, accordingly, the level width broadens. As $E_c$ approaches nearer, the level width rapidly broadens. After $E_c$ comes below the level, it finally disappears. Therefore, every line is terminated at the point where the level width becomes $10^{13}$Hz. As shown in Fig.~\ref{fig3}, the state with $n = 5$ or higher can be broadened by as much as $10^{13}$Hz in an electric field lower than 1MV/cm.

\begin{figure}
\includegraphics{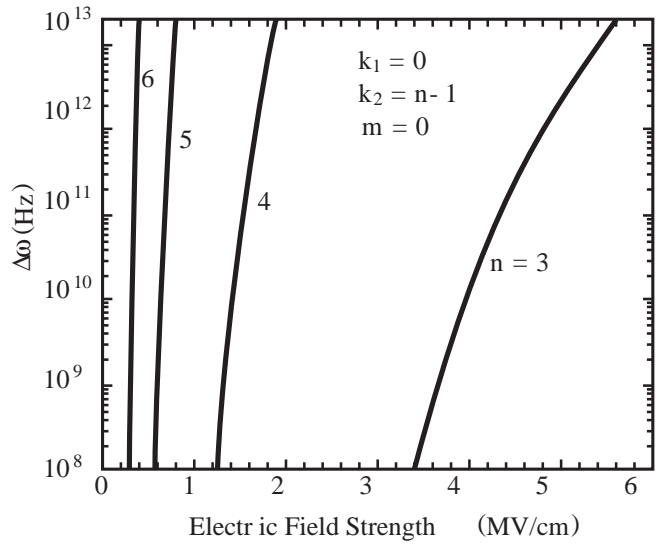}
\caption{Electric field dependence of the ionization probability of Stark states. The level width is shown as the equivalent reciprocal lifetime or the ionization probability.}
\label{fig4}
\end{figure}

The level width is often denoted as the reciprocal lifetime or the ionization probability for a level near the peak of the effective potential barrier. The ionization probability $\Gamma$ (in atomic unit) of a Stark level of a H$^0$ atom is given by the formula \cite{ref13}

\begin{equation}
\Gamma ={1 \over {n^2}}\left( {{4 \over {En^3}}} \right)^{1+\left| m \right|+2k_2}\left\{ {k_2!\left( {k_2+\left| m \right|} \right)!} \right\}^{-1}\times \exp \left\{ {-{2 \over {3n^3E}}+3\left( {n-2k_2-\left| m \right|-1} \right)} \right\}.
\label{eq5}
\end{equation}
$\Gamma$ is converted to frequency as $\Delta \omega = 4.134 \times 10^{16} \Gamma$(Hz). The results of calculations for the Stark states of a H$^0$ atom with $n$ from 3 to 6, $k_1 = 0$, $k_2 = n-1$, and $m = 0$ are shown in Fig.~\ref{fig4}.

The ionization cross section for the combined action of a laser and an electric field is given by $\sigma _i=\sigma _{kn}\eta _i$, where $\sigma _{kn}$ is the transition cross section from a state $k$ to an excited state $n$ by photon absorption, and $\eta _i$ is the ionization probability of the excited state. Since $\eta _i$ is considered to be nearly 1 for a Stark state as broad as $10^{13}$ Hz, $\sigma _i$ is nearly equal to $\sigma _{kn}$.

The transition cross section from a state ($k$) to an upper state ($n$) by absorption of a photon is given by
\begin{equation}
\sigma _{kn}={{g_n} \over {g_k}}{{\lambda _{kn}^2} \over {2\pi }}{{A_{nk}} \over {\Delta \omega _a^{kn}}},
\label{eq6}
\end{equation}
where $g_k$ and $g_n$ are the degeneracies of the states, and equal to 1 in this case, $\lambda _{kn}$ the transition wavelength, $\Delta \omega _a^{kn}$ the line width of the transition, and $A_{nk}$ the Einstein coefficient. The saturation energy density of the transition $k \to n$ is given by
\begin{equation}
\Phi _S^{kn}={{\hbar \omega _{kn}} \over {\sigma _{kn}}},
\label{eq7}
\end{equation}
where $\omega _{kn}$ is the transition frequency. The coherence time of the laser is taken to be sufficiently long to cover the flight time of light from the laser source to the end of the interaction region. However, H$^0$ atoms see the laser as a pulse of the length of the interaction time ($\tau _{int} / \gamma$) in the particle rest frame. Therefore, the necessary energy density of the laser is given by
\begin{equation}
\Phi _{kn}={{\Phi _S^{kn}\gamma } \over {\Delta \omega _a^{kn}\tau _{int }}}.
\label{eq8}
\end{equation}

We are now ready to derive a formula for the necessary laser power density in the laboratory frame. When the laser and H$^0$ beams collide at an angle $\alpha$, the necessary energy density of the laser in the laboratory frame is given by
\begin{equation}
\Phi _{kn}^L={{\Phi _{kn}} \over {\gamma \left( {1+\beta \cos \alpha } \right)}}.
\label{eq9}
\end{equation}
Finally, the necessary laser power density is obtained as
\begin{equation}
I_L={{\Phi _{kn}^L} \over {\tau _{int }}}.
\label{eq10}
\end{equation}

Before evaluating the laser power density, we must define which broad Star state is most suitable. By the selection rule of the dipole transition, $\Delta m = 0$ is allowed for a transition by a laser beam with $\pi$-polarization parallel to the electric field and $\Delta m = \pm1$ by that with $\sigma$-polarization. For a transition with $\Delta m = 0$, the total transition matrix element, or $A_{nk}$ in eq.~(\ref{eq6}), strongly depends upon $k_1$ \cite{ref14, ref15}. For $k_1$ of medium values, $A_{nk}$ decrease to many orders of magnitude smaller than that for $k_1$ near the extreme values: 0 or $n-1$. In addition, the energy level with $k_1 = 0$, $k_2 = n-1$, and $m = 0$ is the lowest of the family of states with the same principal quantum number and accordingly the transition energy is the lowest. Therefore, it seems to be most advantageous to utilize the broad Stark state with $n = 5$, $k_1 = 0$, $k_2 = 4$, and $m = 0$. This state is broadened to $1.7 \times 10^{13}$ Hz in an electric field of 0.8 MV/cm. Under this electric field, the transition energy is shifted to 12.9 eV and, accordingly, the transition wavelength is 96.1 nm in the particle rest frame. 

Although there is no experimental data of the transition cross section in a MV/cm electric field, some theoretical calculations for the transition probability are reported \cite{ref15, ref16}. Here, we use the value $1.38 \times 10^7$ sec$^{-1}$ for $A_{1000,5040}$ given in the Atomic Data and Nuclear Data Table \cite{ref16}. Substituting this value, $\Delta \omega _a = 1.7 \times 10^{13}$ Hz and $\lambda _{1000,5040} = 96.1$nm into eq.(~\ref{eq6}), the cross section is obtained as
\begin{equation}
\sigma _{1000,5040}=1.2\times 10^{-17}cm^2.
\label{eq11}
\end{equation}
Also, the saturation energy density is
\begin{equation}
\Phi _S^{1000,5040}=0.173\left( {{{joule} \over {cm^2}}} \right).
\label{eq12}
\end{equation}

Table~\ref{table1}. shows a result of an evaluation of the relevant parameters for several energies of a H$^0$ beam with the same momentum spread of 0.1\%, and for an interaction length of 30 cm. When the beam energy is 800 MeV, the necessary laser wavelength is in the ultraviolet region. However, as the beam energy increases to 1.0, 1.3 and 2.0 GeV, the laser wavelength changes to violet, blue and yellow. The necessary laser power density is about 4 kW/cm$^2$, and a little higher power density is necessary for a higher beam energy. The magnetic field to cause the necessary Stark effect is below 2 kGauss.

\begin{table}
\caption{Laser Stripping via a Broad Stark State ($n = 5$, $k_1 = 0$, $k_2 = 4$, and $m = 0$),
and Interaction Length = 30cm.
\label{table1}}
\begin{ruledtabular}
\begin{tabular}{l|c|c|c|c|l}
$T$(GeV) & 0.800 & 1.000 & 1.300 & 2.000 & H$^0$ kinetic energy \\ \hline
$\beta$ & 0.842 & 0.875 & 0.903 & 0.948 & \\ \hline
$\gamma$ & 1.853 & 2.066 & 2.386 & 3.132 & \\ \hline
(B$\rho$)(Tm) & 4.881 & 5.657 & 6.778 & 9.288 & Magnetic Rigidity \\ \hline
$\gamma(1+\beta)$ & 3.412 & 3.873 & 4.551 & 6.099 & $\alpha=0$ \\ \hline
(Part. Rest Frame) & & & & & \\ \hline
$\Delta \omega _D$ ($10^{13}$ s$^{-1}$) & 1.65 & 1.71 & 1.81 & 1.86 & Doppler Broadening for $\Delta p/p = 0.001$ \\ \hline
$\tau _{int}^{PRF}$($10^{-9}$s$^{-1}$) & 0.642 & 0.554 & 0.462 & 0.337 & $\tau _{int}^{LF}/ \gamma$ \\ \hline
$\Phi _{1000,5040}$($10^{-6}$ joule/cm$^2$) & 1.59 & 1.84 & 2.20 & 3.02 & $\Phi _S^{1000,5040}=0.173$ (Joul/cm$^2$), $\Delta \omega _a=1.7 \times 10^{13}$ s$^{-1}$ \\ \hline
(Lab. Frame) & & & & & \\ \hline
$\lambda _{1,5}^{LF}$ (nm) & 327.9 & 372.2 & 437.4 & 586.2 & $\lambda _{1,5}^{PRF} = 96.1 $nm \\ \hline
$\Phi _{1000,5040}^{(LF)}$ ($10^{-7}$ joule/cm$^2$) & 4.65 & 4.75 & 4.84 & 4.95 & $\Phi _{1000,5040}/\gamma(1+\beta)$ \\ \hline
$\tau _{int}^{LF}$ ($10^{-9}$s) & 1.19 & 1.14 & 1.10 & 1.06 & $l = 30$ cm \\ \hline
$I^{Laser}$(L.F.) (kW/cm$^2$) & 3.91 & 4.15 & 4.39 & 4.69 & $\Phi _{1000,5040}(LF)/ \tau _{int}^{LF}$ \\ \hline
$B^{Stark}$ (T) & 0.171 & 0.148 & 0.123 & 0.090 & $E = 0.8 \times 10^6$ V/cm \\
\end{tabular}
\end{ruledtabular}
\end{table}

\section{\label{sec:level4} Laser}

With respect to the laser, a narrow-band CW laser is considered to be suitable, because the coherence time of the laser must cover the flight time of light between the laser source to the end of the interaction region, which may be more than several tens of nano-seconds in an actual injection area of the proton ring. Furthermore, the duration of the laser pulse must cover one injection period on the order of 1 msec, while H$^-$ beams are injected into the ring.

At present, the wavelengths of the available lasers are limited. Nd-YAG lasers seem to be the most available, and also their performance is the most reliable. Those have a wavelength of 1,064 nm in the fundamental mode and 532 nm in the second harmonic. For the region near 400 nm, the second harmonic of a Ti-S laser may be useful. The laser wavelength necessary for laser stripping depends on the energy of the H$^-$ beam. We should carefully choose a laser that is reliable and easy to handle.

Nowadays, Fabry-Perot resonators (hereafter denoted as a FP resonator), which have a finesse as high as $10^5$, are realized in some fields of scientific research \cite{ref17, ref18}. Mirrors of such a FP resonator are made of a base material polished to a sub-nanometer roughness and coated by carefully designed multiple layers. The loss rate per one reflection is on the order of one ppm \cite{ref19}. These FP resonators are able to stack a 10kW/cm$^2$ laser beam with a diameter of 10 mm, even when the mirror separation becomes as long as about 6 m. Although the wavelength of a laser that has been actually applied for such a high-finesse FP resonator is limited to 1,064 nm, the fabrication technique is sufficiently applicable to shorter wavelengths. 

The loss rate of photons by reflection and diffraction in one cycle of light in the resonator may be as low as on the order of ppm in a high-finesse FP resonator. However, when we use such a FP resonator for laser stripping, additional photon loss occurs accompanying the absorption of photons to excite H$^0$ atoms to a broad Stark state. The loss rate of photons increases with the intensity of the H$^0$ beam to be stripped. However, the loss rate of photons must be suppressed at a necessary level that depends on the intensity of the available original laser beam injected into the FP resonator to secure the intensity of the laser beam stacked in the FP resonator. In the case of Table~\ref{table1}, where all H$^0$ atoms of the beam are stripped, the laser beam flux is about 4 kW/cm$^2$ and the photon energy is about 3 eV. Accordingly, when the diameter of the laser beam is 1cm, the photon intensity is $1 \times 10^{22}$ photon/s. If the intensity of the H$^0$ beam is $i$ mA, the photon loss due to absorption by the H$^0$ beam occurs at $6.25 i \times 10^{-7}$. Therefore, when the intensity of the source laser is a few hundreds mW, a laser beam with an intensity sufficient to strip an H$^0$ beam of about 100 mA is expected to be stacked in the FP resonator. Thus, a high-finesse FP resonator is considered to be promising for application to a laser stripping system.

\begin{figure}
\includegraphics{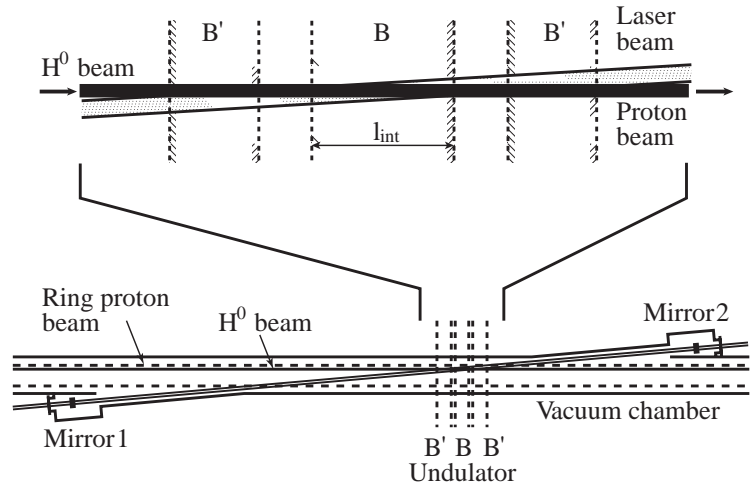}
\caption{Concept of the laser stripping system using a FP resonator and an undulatot. The {\bf mirror 1} and {\bf mirror 2} compose a FP resonator. The H$^0$ beam and the laser beam stacked between {\bf mirror 1} and {\bf mirror 2} collide in a small angle to secure a necessary interaction length {\bf l}$_{int}$. The interaction region is placed in a magnetic field {\bf B} of the central magnet of an undulator to cause H$^0$ atoms necessary Stark effect. The magnetic field {\bf B'} of the outer magnets is set a little different from {\bf B} so that stripping does not occur in these magnets.}
\label{fig5}
\end{figure}

Using such a high-finesse FP resonator, the laser stripping system shown in Fig.~\ref{fig5} may be possible. The H$^0$ and laser beam stacked in the FP resonator collide at a small angle in order to secure the necessary interaction length. An interaction region of about 30 cm long is placed in the central one of three magnets which compose an undulator. The field strength of the central magnet is set so as to cause H$^0$ atoms to experience the necessary Stark effect. The magnetic field strength of the outer magnets is set a little different from that of the central one so that stripping does not occur in these magnets. When the diameters of the H$^0$ and laser beams are taken as 3 mm and 10 mm respectively, the colliding angle is 20 mrad for the interaction length of 30 cm.Since the size of the beam duct is typically 10 cm, the separation of two mirrors of the FP resonator becomes 6 m or longer. Thus, the FP resonator mentioned above is considered sufficiently applicable to this type of laser stripping system.

\section{\label{sec:level5} Tunability of Laser Frequency and Undulator Magnetic Field}

\begin{figure}
\includegraphics{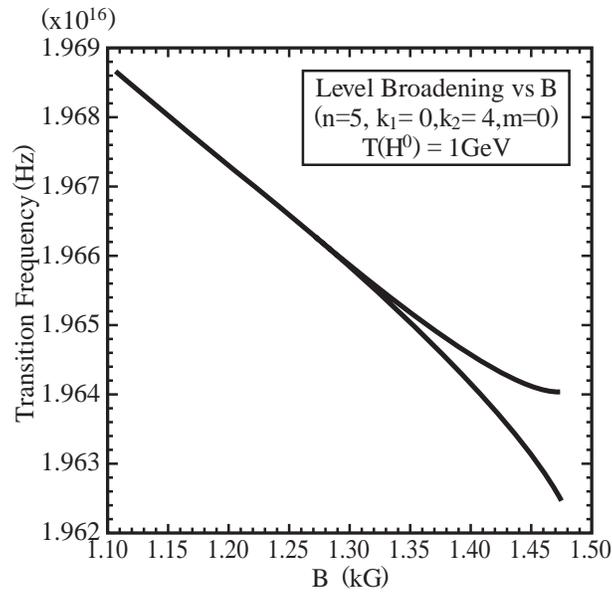}
\caption{Broadening of the level width of the Stark state with $n = 5$, $k_1 = 0$, $k_2 = 4$, and $m = 0$ by the magnetic field strength. The upper (lower) line indicates the transition frequency derived from eq.~(\ref{eq4}) plus (minus) half of the level width derived from eq.~(\ref{eq5}).}
\label{fig6}
\end{figure}

When we put a new method into actual use, it is very important to make clear in advance how to tune all of the necessary parameters. In our case, the tunability of the laser frequency and the magnetic field should be checked. When the kinetic energy of a H$^0$ atom is 1 GeV, the transition frequency of the Stark state with $n = 5$, $k_1 = 0$, $k_2 = 4$, and $m = 0$ shifts and broadens with the magnetic field as is shown in Fig.~\ref{fig6}. From this figure, the operating point is expected to be near to $1.9632 \times 10^{16}$ Hz and 1.47 kGauss. However, we must probably search for the best operating point, because not only the calculation, but also the caliblation of laser frequency and the magnetic field, often includes some errors. In this case, since the level width is as large as $10^{13}$ Hz near the operating point, such a search procedure may be considered. The laser frequency of the FP resonator should be changeable in steps of a few times $10^{12}$ Hz over a frequency range that well includes the operating point. The magnetic field of the undulator should be changeable by steps of 10 or 20 Gauss up to a sufficiently high field. Then, we scan the laser frequency and the magnetic field, while observing the yield of protons downstream of the interaction region. The peak of the proton yield indicates the best operating point.

\section{\label{sec:level6} Emittance Growth due to Stripping in the undulator}

\begin{figure}
\includegraphics{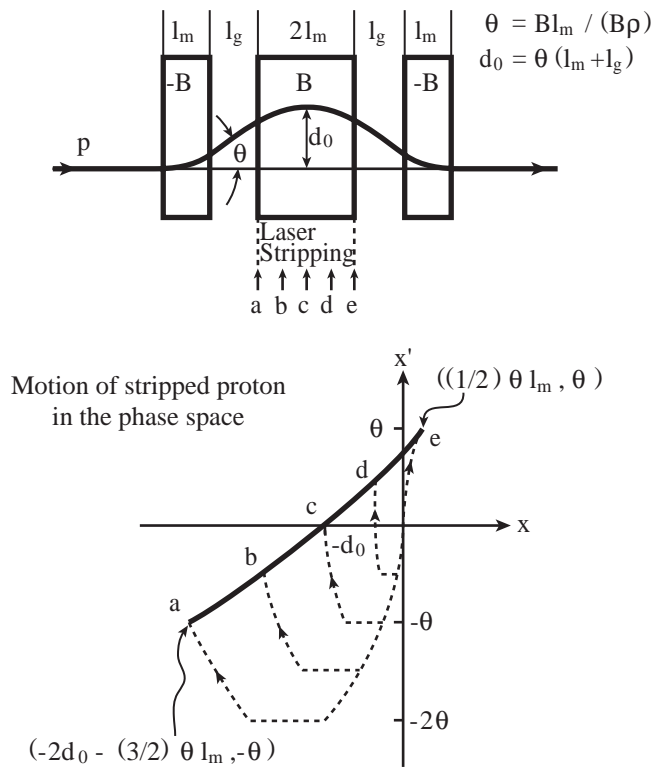}
\caption{(Upper): Deflection and displacement in the undulator magnet. (B$\rho$) is the magnetic rigidity of the proton beam. H$^0$ and laser beam collide in the magnetic field of the central of the undulator magnet and accordingly stripping takes place there. (Lower): Motion of the formed protons in the phase space. Protons formed by laser stripping at the point {\bf a} through {\bf e} in the central magnet walk along respective dotted line from the origin to the point {\bf a} through {\bf e} int phase space while protons move from the point they have been stripped to the end of the undulator magnet.}
\label{fig7}
\end{figure}

Another important item to be checked is the emittance growth due to the stripping process. Here, we consider the case where stripping takes place in the central magnet of the undulator, as is shown in Fig.~\ref{fig7}(Upper). For the sake of simplicity, the magnetic fields of the outer magnets are taken to be the same as that of the central magnet. Then, the protons circulating in the ring take the orbit shown by the thick line. The maximum deflection ($\theta$) and the maximum displacement (d$_0$) are given by the equations shown on the right side of the figure.

We consider that laser stripping occurs only in the central magnet. The motion of protons stripped at various points of the magnet is shown in Fig. ~\ref{fig7}(Lower). While a proton stripped at point {\bf a} in the central magnet moves and reaches the end of the undulator, it walks along the dotted line from the origin, and reaches point {\bf a} in phase space. In the same manner, the protons formed at points {\bf b} through {\bf e} in the central magnet walk along the respective dotted line from the origin to points {\bf b} through {\bf e} in phase space. Therefore, at the end of the undulator, protons just formed by stripping distribute along the thick line in phase space. Thus, ptotons have a deflection spread of $2\theta$, and a displacement spread of $2($d$_0+\theta l_m)$. For example, when 1 GeV H$^0$ beams are stripped in an undulator magnet with B = 0.148 T, $l_m = 15$ cm and $l_g = 10$ cm, $\theta$  and $($d$_0+\theta l_m)$ are 3.92 mrad and 1.57 mm, respectively.

These spreads of the deflection and displacement increase the emittance of the formed proton beam by several times the original emittance of the H$^0$ beam which is typically 1 $\pi$mmmrad. However, such an emittance growth is not considered to be very serious in a situation where we need to form a ring beam with an emittance larger than 100 $\pi$mmmrad by sophisticated phase-space painting. The emittance growth accompanying laser stripping should be taken into account as a part of the phase-space painting.

\section{\label{sec:level7} Conclusion}

Resonant photo-ionization via a broad Stark state formed at energies near the peak of the effective potential barrier, called here laser stripping via a broad Stark state, turned out to be effective not only to cover the spread of the transition-frequency distribution due to the momentum spread of the H$^0$ (or H$^-$) beam, but also to avoid saturation of the stripping efficiency to one half by the competition of pumping up and down. As a result, the stripping efficiency is expected to reach 100\%.

For a 1 GeV H$^0$ beam, the magnetic field necessary to broaden the Stark state of a H$^0$ atom with $n = 5$, $k_1 = 0$, $k_2 = 4$, and $m = 0$ as broad as $10^{13}$ Hz is estimated to be 1.45 kGauss, and the laser power density necessary to complete stripping in an interaction length of 30 cm is estimated to be 4.15 kW/cm$^2$. Today, high power FP resonators are available which have finesses near $10^5$ and lengths of 6m or longer, and are able to stack a 10 kW/cm$^2$ light beam with a diameter of 10 mm. These FP resonators are considered to be sufficiently applicable for the laser stripping systems installed in actual accelerators. 

Because the level width of the Stark state is very broad, tuning of the laser frequency and the undulator magnetic field is rather easy. The emittance growth accompanying stripping in a magnetic field is expected to be several times the emittance of the H$^0$ beam, but is insignificant in a situation where a ring beam with an emittance larger than 100 $\pi$mmmrad is formed from the H$^-$ beam by sophisticated phase-space painting. This emittance growth should be taken into account as a part of beam painting.

Thus, a laser stripping system using a high-finesse Fabry-Perot resonator and an undulator magnet is considered to be sufficiently applicable to actual high-intensity proton rings as a substitute for the conventional injection system using a solid foil.


\end{document}